%% file: ANA-HDBS-2018-33-PAPER.tex
\newcommand*{\ATLASLATEXPATH}{}
\documentclass[PAPER, atlasdraft=true, texlive=2016, UKenglish, cernpreprint]{\ATLASLATEXPATH atlasdoc}
 
\usepackage{\ATLASLATEXPATH atlaspackage}
\usepackage{\ATLASLATEXPATH atlasbiblatex}
 
\usepackage[BSM=true,jetetmiss=true,process=true]{\ATLASLATEXPATH atlasphysics}
 
\graphicspath{{logos/}{figures/}}
 
\usepackage{ANA-HDBS-2018-33-PAPER-defs}
\usepackage{tabularx}

\AtlasTitle{Search for non-resonant Higgs boson pair production in the $bb \ell \nu \ell \nu$ final state
with the ATLAS detector in $pp$ collisions at $\sqrt{s} = 13$ TeV}

\PreprintIdNumber{CERN-EP-2019-143}
 
\AtlasAbstract{
A search for non-resonant Higgs boson pair production, as predicted by the Standard Model, is presented,
where one of the Higgs bosons decays via the $H\rightarrow bb$ channel and the other via one of
the $H \rightarrow WW^*/ZZ^*/\tau\tau$ channels.
The analysis selection requires events to have at least two $b$-tagged jets and exactly two leptons (electrons or muons)
with opposite electric charge in the final state.
Candidate events consistent with Higgs boson pair production are selected using a multi-class neural network discriminant.
The analysis uses 139~fb$^{-1}$ of $pp$ collision data recorded at a centre-of-mass energy of 13~TeV by the
ATLAS detector at the Large Hadron Collider.
An observed (expected) upper limit of 1.2 ($0.9^{+0.4}_{-0.3}$) pb
is set on the non-resonant Higgs boson pair production cross-section at 95\% confidence level,
which is equivalent to 40 ($29^{+14}_{-9}$) times the value predicted in the Standard Model.
}
 
\author{The ATLAS Collaboration}
 
\AtlasRefCode{HDBS-2018-33}

\AtlasJournal{Phys. Lett. B 801 (2020) 135145}
\AtlasDOI{10.1016/j.physletb.2019.135145}

\hypersetup{pdftitle={ATLAS document},pdfauthor={The ATLAS Collaboration}}
 
\begin{document}
 
\maketitle

% The next lines are included from the .//sections/intro.tex input file
\section{Introduction}
\label{sec:intro}

\begin{figure}[!htb]
\begin{center}
\includegraphics[height=0.18\textwidth]{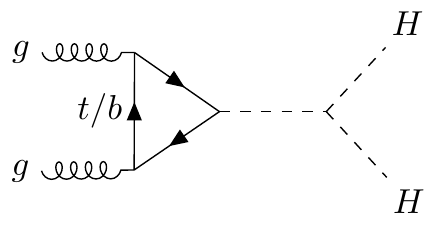}\qquad
\includegraphics[height=0.18\textwidth]{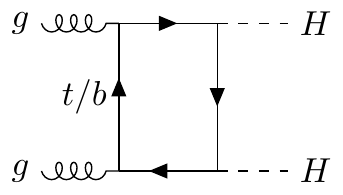}
\caption{
Feynman diagrams for leading order ggF production of Higgs boson pairs:
the `triangle diagram' sensitive to the Higgs boson self-coupling on the left and
the `box diagram' on the right.
}
\label{fig:hh_diagrams}
\end{center}
\end{figure}

In 2012, the ATLAS and CMS Collaborations reported the observation of a new particle
in the search for the Standard Model (SM) Higgs boson (\Hboson) \cite{HIGG-2012-27,CMS-HIG-12-028}.
So far, measurements of the spin and couplings of the new particle are consistent with
those predicted by the Brout--Englert--Higgs (BEH) mechanism of the SM \cite{HIGG-2013-17,HIGG-2015-06,HIGG-2018-13,HIGG-2018-04,CMS-HIG-15-002,CMS-HIG-17-035,CMS-HIG-18-016,Sirunyan:2019twz,CMS-HIG-17-034, CMS-HIG-17-031}.
The SM predicts non-resonant production of Higgs boson pairs (\HH) in proton--proton ($pp$) collisions, referred to as \textit{non-resonant \HH production},
with the dominant production modes at the LHC proceeding via the gluon--gluon fusion (ggF) process.
The ggF process has two leading order contributions: the first corresponds to the so-called `triangle diagram', which includes the Higgs boson self-coupling,
and the second is the so-called `box diagram', which includes a heavy-quark loop with
two fermion--fermion--Higgs ($ffH$) vertices. 
These two amplitudes interfere destructively, resulting in a low cross-section of only $31.05\pm1.90$ fb for the
ggF \hh production mode, computed at next-to-next-to-leading order (NNLO) and including
finite top-quark mass effects \cite{deFlorian:2016spz,Dawson:1998py,Borowka:2016ehy,Baglio:2018lrj,deFlorian:2013jea,Shao:2013bz,deFlorian:2015moa,Grazzini:2018bsd}.
Feynman diagrams illustrating these two contributions are shown in Figure \ref{fig:hh_diagrams}.
The measurement of non-resonant \HH production at the LHC stands as an important test of the BEH mechanism.
In many beyond-the-SM (BSM) theories, \HH production can be enhanced by modifying the Higgs boson self-coupling, \lhhh,
or the top-quark Yukawa coupling, $y_t$, and/or by introducing new contact interactions between two top-quarks or gluons and two Higgs bosons or introducing production mechanisms via intermediate BSM particles \cite{Branco:2011iw,Azatov:2015oxa,Huang:2017nnw}.

The ATLAS and CMS Collaborations have performed searches for non-resonant \HH production in a variety of final states at 13~\TeV~
\cite{HIGG-2016-15,HIGG-2016-16,HIGG-2016-20,HIGG-2016-24,HIGG-2016-27,EXOT-2016-31,Sirunyan:2017djm,CMS-HIG-17-006,CMS-HIG-17-008,CMS-HIG-17-017}. 
No significant excess of events beyond SM expectations is observed in these searches, with the ATLAS and CMS data-analyses setting
observed (expected) limits on non-resonant \hh production to be no larger than 6.9 (10.0) and 22.2 (12.8) times the predicted rate in the SM, respectively \cite{Aad:2019uzh, CMS-HIG-17-030}.

This Letter describes a search for non-resonant \HH production in the \bblvlv final state, where $\ell$ refers
to a lepton (either an electron or a muon),
using 13~\TeV~\pp collision data collected with the ATLAS detector during 2015--2018 and corresponding to
a total integrated luminosity of 139~\ifb.
The analysis uses machine-learning techniques based on feedforward neural network architectures \cite{Goodfellow-et-al-2016}
to construct an event-level classifier
trained to distinguish between the $\hh$ signal and SM backgrounds.
Analyses searching for non-resonant \HH production via similar decay channels were performed previously
in the single-lepton final state by ATLAS in searches for $\hh \rightarrow \bbWW$ \cite{HIGG-2016-27}
and in the dilepton channel by CMS in searches for $\hh \rightarrow \bbWW/\bbZZ$ \cite{CMS-HIG-17-006}.
% End of text imported from the .//sections/intro.tex input file
% The next lines are included from the .//sections/detector.tex input file
\section{ATLAS detector}
\label{sec:detector}
 
The ATLAS detector \cite{PERF-2007-01,ATLAS-TDR-2010-19,PIX-2018-001} is a general-purpose particle detector with forward--backward
symmetric cylindrical geometry.\footnote{
ATLAS uses a right-handed coordinate system with its origin at the nominal interaction point
(IP) in the centre of the detector and the $z$-axis along the beam pipe. The $x$-axis points
from the IP to the centre of the LHC ring, and the $y$-axis points upwards. Cylindrical coordinates
$(r,\phi)$ are used in the transverse plane, $\phi$ being the azimuthal angle around the $z$-axis.
The pseudorapidity is defined in terms of the polar angle $\theta$ as $\eta = -\ln \tan(\theta/2)$.
The angular distance is measured in units of $\Delta R \equiv \sqrt{(\Delta\eta)^{2} + (\Delta\phi)^{2}}$.}
It includes an inner tracking detector (ID), immersed in an axial magnetic field,
which provides precision tracking of charged particles over the range of $|\eta|<2.5$.
Calorimeter systems with either liquid argon or scintillator tiles as the active medium provide energy
measurements over the range of $|\eta|<4.9$.
The muon spectrometer (MS) is positioned outside the calorimeters and includes three air-core toroidal magnets.
The MS is composed of several types of muon detectors
which provide trigger and high-precision tracking capabilities for $|\eta|<2.4$
and $|\eta|<2.7$, respectively. A hardware-based trigger followed by a software-based trigger reduce the recorded event rate to an average of 1~kHz \cite{TRIG-2016-01}.
% End of text imported from the .//sections/detector.tex input file
% The next lines are included from the .//sections/datamc.tex input file
\section{Dataset and simulated events}
\label{sec:datamc}
 
The data used for this search were collected in \pp collisions at the LHC with a centre-of-mass energy of 13~\tev.
Only those data collected during stable LHC beam conditions and with all ATLAS detector
subsystems fully operational are used, and correspond to an integrated luminosity of
139~\ifb.
The selection of candidate events with oppositely charged leptons is based
on a combination of single-lepton and dilepton triggers.\footnote{Distinct sets of single-lepton triggers are used for electrons and muons. Dilepton triggers require either two electrons, two muons, or one electron and one muon.}
The use of a given trigger depends on the flavour and the transverse momenta (\pt) of the two
(\pt-ordered) leptons in the event, and on the data-taking period.
Single-lepton triggers with \pt thresholds
between $22$ and $28~\gev$ are given priority over dilepton triggers. The criteria of the dilepton triggers are checked only 
if no single-lepton trigger criteria are met and have \pt thresholds as low as 19 (10)$~\gev$ for the leading (subleading) lepton.
At least one reconstructed lepton (or lepton pair) has to match a corresponding trigger
object, in which case their offline \pt must be higher than the trigger threshold by
at least $2$ GeV, in order to be on the efficiency plateau of the corresponding trigger.
 
Monte Carlo (MC) simulation \cite{SOFT-2010-01} is used to model the signal processes
and in the estimation of SM background processes.
A \textsc{GEANT4} \cite{GEANT4} simulation of the ATLAS detector was used for the background processes.
The signal MC samples were processed with a fast simulation that relies on a
parameterisation of the calorimeter response \cite{SOFT-2010-01} and on \textsc{GEANT4} for the tracking detectors.
Simulated events are reconstructed using the same algorithms as used for data and include the
effects of multiple \pp interactions in the same or neighbouring bunch crossings,
collectively referred to as \pileup.
The simulation of \pileup collisions
was performed with \textsc{Pythia} 8.186 \cite{Sjostrand:2007gs} using the ATLAS A3 set of tuned parameters \cite{ATL-PHYS-PUB-2016-017}
and the NNPDF2.3LO parton distribution function (PDF) set \cite{Ball:2012cx}.
Simulated events were reweighted to match the distribution of \pileup interactions in data. The average amount of \pileup in the data collected during 2015--2018 was $33.7$.
 
The signal processes with ggF-initiated non-resonant \HH production in the \bblvlv
final state were generated with an effective Lagrangian in the infinite top-quark mass
approximation.
The generated signal events were reweighted with form factors that take into account
the finite mass of the top-quark \cite{CP3Paper,Degrassi_Ramona}.
SM background processes were simulated using different MC event generators. The MC matrix element (ME) event generators and PDF sets,
the parton showering (PS) and the underlying event (UE) modelling, UE tuned parameters (tune),
and the accuracy of the theoretical cross-sections used to normalise the simulated processes
are summarised in Table \ref{tab:data_mc}.
Each SM background process is normalised to the best available respective
theoretical cross-section. The mass of the Higgs boson was set to 125~\gev\ for all signal and background processes.
The \HH branching fractions (BF) predicted by the SM \cite{deFlorian:2016spz} are used for all Higgs boson decays.
\textsc{MadSpin} \cite{Artoisenet:2012st}
was used to model top-quark spin correlations and \textsc{EvtGen} \cite{Lange:2001uf} was used to model properties of $b$- and $c$-hadron decays
for processes using \textsc{Pythia} and for the signal processes.

SM top-quark pair production (\ttbar) and the production of single top-quarks in association with $W$ bosons (\wt) contribute with significant
background contamination in the \bblvlv final state.
At next-to-leading-order (NLO) accuracy, there exists non-trivial interference between
these two processes that may be enhanced in phase-space regions wherein there are high fractions
of \wt events \cite{TOPQ-2017-05}.
Two schemes are typically used to remove the overlap between these two processes: the so-called
diagram removal (DR) and diagram subtraction (DS) schemes \cite{Frixione:2008yi}; the former
is used in the present analysis to remove the overlapping events and the latter is used to evaluate the
systematic uncertainty in corresponding background event yields.
Because of these effects, the sum of the simulated \ttbar and \wt processes is considered
as a single background process and referred to as the `\bkgtop' process in what follows.
 
% The next lines are included from the .//tables/tab_data_mc.tex input file
\begin{table}[!htb]
\scriptsize
\begin{center}
\caption{List of the ME generators and PS/UE modelling algorithms used in the simulation. Alternative generators and PS/UE models, used to estimate systematic uncertainties, are shown in parentheses. The PDF sets, tunes, and the perturbative QCD highest-order accuracy (leading-order, LO; next-to-leading-order, NLO; next-to-next-to-leading-order, NNLO; next-to-next-to-leading-logarithm, NNLL) used for the normalisation of the samples are also included. The top-quark mass is set to 172.5~\GeV.}
\label{tab:data_mc}
\scalebox{0.78}{ \begin{tabular}{ l | l | l | l | l | l }
\toprule
\hline
Process & ME generator & ME PDF & PS/UE model & UE tune & Prediction order for \\
& (alternative) & & (alternative) & & total cross-section\\
\hline
\ttbar~\cite{ATL-PHYS-PUB-2018-009,Campbell:2014kua} & \powhegbox\textsc{ v2} \cite{Alioli:2010xd,Frixione:2007vw} & NNPDF3.0NLO \cite{Ball:2014uwa} & \pythia~8.230 \cite{Sjostrand:2014zea} & A14 \cite{ATL-PHYS-PUB-2014-021} & NNLO + NNLL \cite{Beneke:2011mq,Cacciari:2011hy,Baernreuther:2012ws,Czakon:2012zr,Czakon:2012pz,Czakon:2011xx,Czakon:2013goa} \\
& (\mgfive) & & (\herwig~7.0.4) & (H7-MMHT14) & \\
Single-top $s$-channel, \wt \cite{ATL-PHYS-PUB-2018-009,Alioli:2009je,Re:2010bp} & \textsc{Powheg-Box} & NNPDF3.0NLO & \pythia~8.230 & A14 & NLO + NNLL \cite{Kidonakis:2010ux,Kidonakis:2010tc} \\
& (\mgfive) & & (\herwig~7.0.4) & (H7-MMHT14) & \\
Single-top $t$-channel \cite{ATL-PHYS-PUB-2018-009,Alioli:2009je} & \powhegbox, \textsc{MadSpin} \cite{Artoisenet:2012st} & NNPDF3.04fNLO & \pythia~8.230 & A14 & NLO + NNLL \cite{Kidonakis:2011wy} \\
& (\mgfive) & & (\herwig~7.0.4) & (H7-MMHT14) & \\
$W,Z/\gamma^*+\mathrm{jets}$ \cite{ATL-PHYS-PUB-2017-006} & \sherpa~2.2.1 \cite{Gleisberg:2008ta,Bothmann:2019yzt} & NNPDF3.0NNLO & \sherpa~2.2.1 & \sherpa~default & NLO(LO) $\leq$2(4) partons \cite{Catani:2009sm,Gleisberg:2008fv,Cascioli:2011va,Schumann:2007mg,Hoeche:2012yf} \\
($Z/\gamma^*+\mathrm{jets}$) & (\mgfive) & & (\pythia~8.230) & (A14) & \\
Diboson ($WW,~WZ,~ZZ$) \cite{ATL-PHYS-PUB-2017-005} & \sherpa~2.2.2 & NNPDF3.0NNLO & \sherpa~2.2.2 & \sherpa~default & NLO(LO) $\leq$1(3) partons \cite{Gleisberg:2008fv,Cascioli:2011va,Schumann:2007mg,Hoeche:2012yf} \\
$\ttbar W$,$\ttbar Z$ \cite{ATL-PHYS-PUB-2016-005} & \mgfive \cite{Alwall:2014hca} & NNPDF3.0NLO & \pythia~8.210 & A14 & NLO \cite{Campbell:2012dh,Garzelli:2012bn} \\
$\ttbar H$ \cite{ATL-PHYS-PUB-2016-005} & \mgfive & NNPDF3.0NLO & \pythia~8.210 & A14 & NLO \cite{Frixione:2014qaa,Frixione:2015zaa} \\
$WH,~ZH$ \cite{ATL-PHYS-PUB-2018-035} & \pythia~8.186 \cite{Sjostrand:2007gs} & NNPDF2.3LO \cite{Ball:2012cx} & \pythia~8.186 & A14 & NNLO QCD + NLO EW \cite{Ciccolini:2003jy,Brein:2003wg,Ferrera:2011bk,Brein:2011vx,Ferrera:2013yga,Ferrera:2014lca,Campbell:2016jau} \\
ggF $H$ \cite{ATL-PHYS-PUB-2016-006} & \powhegbox\textsc{ v2}~NNLOPS \cite{Hamilton:2013fea}& CT10 \cite{Lai:2010vv} & \pythia~8.212 & AZNLO \cite{STDM-2012-23} & NNNLO QCD + NLO EW \cite{Anastasiou:2016cez}\\
\hline
SM $\hh\rightarrow\bblvlv$ \cite{ATL-PHYS-PUB-2019-007} & \mgfive~2.6.2 & CT10 & \herwig~7.0.4 \cite{Bellm:2015jjp} & H7-MMHT14 \cite{Harland-Lang:2014zoa} & NNLO \cite{Dawson:1998py,Borowka:2016ehy,Baglio:2018lrj,deFlorian:2013jea,Shao:2013bz,deFlorian:2015moa,Grazzini:2018bsd} \\ 
\hline
\bottomrule
\end{tabular}}
\end{center}
\end{table}
% End of text imported from the .//tables/tab_data_mc.tex input file
% End of text imported from the .//sections/datamc.tex input file
% The next lines are included from the .//sections/objects.tex input file
\section{Event selection and object definitions}
\label{sec:objects}
 
Selected events are required to have at least one \pp interaction vertex reconstructed
from at least two ID tracks with $\pt>0.4~\gev$. The primary vertex for each event is
defined as the vertex with the highest $\sum\left(\pt\right)^2$ of associated ID tracks \cite{ATL-PHYS-PUB-2015-026}.
Events that contain at least one jet arising from
non-collision sources or detector noise are rejected by a set of quality
criteria \cite{ATLAS-CONF-2015-029}.
 
\textit{Loose} and \textit{signal} criteria are defined in order to select reconstructed lepton
and jet candidates, where the latter is a subset of the former. Compared to the loose objects,
the signal objects are required to satisfy tighter identification or quality criteria
that are designed to suppress background contributions.
Reconstructed loose (signal) electrons are required to satisfy the `Loose' (`Tight')
likelihood identification criteria \cite{LatestEgamma}. Loose electrons are required
to have $\pt>10~\gev$ and to be within $|\eta|<2.47$. In addition, signal electrons are
required to be outside the range $1.37<|\eta|<1.52$, which corresponds to the transition
regions between the barrel and endcaps of the electromagnetic calorimeters.
In order to reduce background contributions from jets misidentified as electrons, signal electrons
are required to be isolated according to the `Gradient' selection criteria \cite{LatestEgamma}.
Reconstructed loose and signal muon candidates are required to have $\pt>10~\gev$, to be
within $|\eta|<2.4$, and to satisfy the `Medium' identification criteria \cite{PERF-2015-10}.
Additionally, signal muons are required to be isolated according
to the `FixedCutLoose' selection criteria \cite{PERF-2015-10}.
Signal electron (muon) candidates are required to originate from the primary vertex by demanding that the
significance of the transverse impact parameter, defined as the absolute value of the track transverse
impact parameter, $d_0$, measured relative to the primary vertex, divided by its uncertainty,
$\sigma_{d_0}$, satisfy $|d_0|/\sigma_{d_0} < 5$ $(3)$. The difference $\Delta z_0$ between
the value of the $z$ coordinate of the point on the track at which $d_0$ is defined and the
longitudinal position of the primary vertex is required to satisfy $|\Delta z_0 \times \sin \theta| < 0.5$ mm,
where $\theta$ is the polar angle of the track with respect to the $z$-axis.
 
Jets are reconstructed from topological clusters of energy deposits in the calorimeters
\cite{PERF-2014-07} using the anti-$k_t$ algorithm \cite{Cacciari:2008gp,Fastjet} with a radius parameter of $R=0.4$ and calibrated as
described in Ref. \cite{PERF-2016-04}.
Candidate loose jets are required to have $\pt>20~\gev$.
Signal jets are required to have $|\eta|<2.8$ and must satisfy \pileup suppression requirements
based on the output of a multivariate classifier \cite{PERF-2014-03}, which identifies
jets consistent with a primary vertex in the region $|\eta|<2.4$ and $\pt<120~\gev$.
The MV2c10 multivariate algorithm \cite{FTAG-2018ttbar} is used to identify
jets containing $b$-hadrons ($b$-tagged jets). An MV2c10 working point with a $b$-tagging efficiency of $70\%$, estimated from
simulated $\ttbar$ events \cite{PERF-2016-05}, is used.
The $b$-tagged jets must have $\pt>20~\gev$
and $|\eta|<2.5$.
The momentum of $b$-tagged jets is adjusted using the muon-in-jet correction, as
described in Ref. \cite{HIGG-2018-04}, by
accounting for momentum losses due to muons originating from in-flight semileptonic $b$-hadron decays occurring
within the $b$-tagged jet.
 
The missing transverse momentum \ptmiss, the magnitude of which is denoted by \met, is constructed
from the negative vectorial sum of the transverse momenta of calibrated loose objects in the event.
An additional term is included to account for the energy of ID tracks that are matched
to the primary vertex in the event but not to any of the selected loose objects \cite{PERF-2016-07}.
 
To avoid double-counting, loose objects are subject to the overlap removal procedure defined as follows.
If a reconstructed electron and muon share a track in the ID, the electron
is removed. However, if the muon sharing the track with the electron is calorimeter-tagged,\footnote{A calorimeter-tagged
muon has only a reconstructed track in the ID matched to energy deposits in the calorimeter compatible
with a minimum ionising particle, but no corresponding track segment in the MS.}
then the muon is removed instead of the electron.
If a jet and an electron are reconstructed within $\Delta R = 0.2$ of each other,
then the jet is removed.
If a jet and a muon are within $\Delta R = 0.2$ of each other, and the jet has less than three tracks or carries less than $50\%$ of the muon
\pt, then the jet is removed; otherwise, the muon is removed.
Electrons or muons separated from the remaining jets by $\Delta R < 0.4$ are removed.
 
The analysis selects candidate events with exactly two oppositely charged signal leptons, electrons or muons,
and at least two signal $b$-tagged jets. To enhance sensitivity to the signal process
and to maximise rejection of the expected SM backgrounds, the analysis uses a
multivariate approach to select signal events.
% End of text imported from the .//sections/objects.tex input file
% The next lines are included from the .//sections/analysis.tex input file
\section{Analysis strategy}
\label{sec:analysis}
 
The analysis relies on the use of a multivariate discriminant designed to select candidate events consistent
with non-resonant \HH production. Section \ref{sec:nn_desc} describes the architecture and the training of the deep
neural network (DNN) classifier from which the discriminant is constructed. Section \ref{sec:signal_selection} describes the
signal region selection criteria. Section \ref{sec:background_estimation} describes the final background estimation procedure.
 
% The next lines are included from the .//sections/deep_learning_approach.tex input file
\subsection{Deep learning approach to target $\hh$}
\label{sec:nn_desc}
 
The discriminant uses the outputs of a DNN classifier that is
built using the \textsc{Keras} library with \textsc{Tensorflow} as a backend \cite{chollet2015keras,tensorflow2015-whitepaper}
and uses the \textsc{lwtnn} library \cite{daniel_hay_guest_2019_2583131} to interface with the analysis
software infrastructure of the ATLAS experiment.
The sample of events used for training is composed of equal numbers of events from the signal
and each of the dominant background processes: Top (as defined in Section \ref{sec:datamc}),
$Z/\gamma^* \rightarrow \ell\ell$ (Z-$\ell\ell$), and $Z/\gamma^* \rightarrow \tau\tau$ (Z-$\tau\tau$) production.
The signal sample used in the training of the classifier contains only the
$\hh \rightarrow \bbWW$ component due to its larger BF relative to the $\hh \rightarrow \bbtt$ and
$\hh \rightarrow \bbZZ$ components. However, the sum of all three signal components is evaluated as the signal when performing the statistical analysis.
Additionally, all processes that make up the training sample ($\hh \rightarrow \bbWW$, \bkgtop, Z-$\ell\ell$, and Z-$\tau\tau$)
have the same weight during the training of the classifier.
The training sample is composed of simulated candidate events with $\mll>20~\gev$ and having one or more $b$-tagged jets, where events
with exactly one $b$-tagged jet are included to increase the number of events available for training.
For the training events with exactly one $b$-tagged jet, each observable that requires
at least two $b$-tagged jets is set to its mean value as computed with the
full set of training events that contain at least two $b$-tagged jets.
Observables that require two $b$-tagged jets are defined using the leading two
$b$-tagged jets.
The classifier contains two fully connected hidden layers each with 250 nodes. Rectified linear unit (ReLU)
activations are used for each layer \cite{Nair:2010:RLU:3104322.3104425}. In order to improve
the robustness of the training and to reduce effects due to overtraining, there is a dropout layer that randomly drops 50\% of the nodes between the two fully connected layers during training \cite{JMLR-Dropout}.
The classifier produces four outputs
that are passed through a softmax activation,
constraining their sum to one \cite{Goodfellow-et-al-2016}. The resulting four outputs, each constrained to values between 0 and 1,
are referred to as
$p_{i}$ ($i \in \{\hh, \text{Top}, \text{Z-}\ell\ell, \text{Z-}\tau\tau\}$). Values of $p_{i}$
nearer to 1 indicate that the event likely belongs to class $i$ and values nearer
to 0 indicate otherwise.
The main discriminant in the analysis, $\dhh$, is constructed from the four $p_{i}$ and is defined
as $\dhh = \ln \left[ \phh / \left( \ptop + \pzsf + \pztt \right) \right]$.
 
The $\HH \rightarrow \bblvlv$ signal events are characterised by two distinct `Higgs hemispheres'.
One hemisphere contains the two $b$-tagged jets from the $H \rightarrow bb$ decay and it is
typically opposite in the transverse plane to the second hemisphere that contains the two leptons
and \met from the $H \rightarrow WW^*/ZZ^*/\tau\tau$ decay.
The final-state objects in the SM backgrounds, the \bkgtop process in particular, are distributed
more uniformly within the event and they typically do not exhibit the same opposite hemispheres
topology as the \hh signal.
These Higgs hemispheres thus provide a topological criterion that distinguishes the signal
from the background and motivates the choice of input observables that are provided to
the classifier.
Thirty-five such variables are provided as inputs to the classifier, ranging from momentum components
of the visible final-state objects to observables using event-wide information, and are constructed
using only calibrated final state objects that have well-understood uncertainties (Section \ref{sec:systematics}).
A complete list is provided in Table \ref{tab:nn_inputs}.
The event-wide input observables are sensitive to the presence of Higgs hemispheres
in the signal and are largely angular in nature or take advantage of the fact that the final
state objects from each of the Higgs bosons in the signal tend to be near to each other.
The observables \HTRatio and \mtbb are non-standard high-level observables that are not straightforward functions of the momenta
of the final-state objects.
By construction, \HTRatio can take values between zero and one; it peaks near one for signal
and is more broadly distributed for background.
The \mtbb observable is defined similarly to the \cmsmtbb observable
in Ref. \cite{CMS-TOP-15-008} but does not include the final-state leptons.
As discussed in Ref. \cite{CMS-TOP-15-008}, for the \bkgtop backgrounds \mtbb generally
has values below the mass of the top-quark due to kinematic constraints while for the
$Z/\gamma^*$ processes, which have little-to-no \met, \mtbb is typically below $45$ GeV.
The use of dropout regularisation during the training of the classifier allows it
to more effectively use the information contained in the full set of inputs presented in Table \ref{tab:nn_inputs}
by reducing its susceptibility to overtraining effects that may otherwise appear as a result of using such
an extended input feature space in the case where no such regularisation is performed.
To verify this, the performance of the classifier was checked using an independent sample of events not used in the
training of the classifier and was found to be compatible to its performance when presented with those
of the training sample.
 
% The next lines are included from the .//tables/tab_nn_inputs.tex input file
\begin{table}[!htb]
\begin{center}
\begin{footnotesize}
\caption{
Description of the variables used as inputs to the DNN classifier.
}
\label{tab:nn_inputs}
\begin{tabularx}{\textwidth}{l |p{13cm}l}
\toprule
\hline
($p_{T}$, $\eta$, $\phi$) & $p_{T}$, $\eta$, and $\phi$ of the leptons, leading two jets (not necessarily $b$-tagged), and leading two $b$-tagged jets \\
Dilepton flavour & Whether the event is composed of two electrons, two muons, or one of each (encoded as 3 booleans)\\
$\Delta R_{\ell\ell}$, $|\Delta \phi_{\ell\ell}|$   & $\Delta R$ and magnitude of the $\Delta \phi$ between the two leptons \\
$m_{\ell\ell}$, $p_{T}^{\ell\ell}$  & Invariant mass and the transverse momentum of the dilepton system \\
$\met$, $\met$-$\phi$ & Magnitude of the missing transverse momentum vector and its $\phi$ component \\
$|\Delta \phi (\ptmiss, \ptvec{\ell}{\ell})|$ & Magnitude of the $\Delta \phi$ between the \ptmiss and the transverse momentum of the dilepton system \\
$|\ptmiss + \ptvec{\ell}{\ell}|$ & Magnitude of the vector sum of the \ptmiss and the transverse momentum of the dilepton system \\
Jet multiplicities & Numbers of $b$-tagged and non-$b$-tagged jets \\
$|\Delta \phi_{bb}|$ & Magnitude of the $\Delta \phi$ between the leading two $b$-tagged jets \\ [0.08cm]
$\mtbb$ & \mttwo \cite{MT2-Glamour} using the leading two $b$-tagged jets as the visible inputs and \ptmiss as invisible input\\ [0.1cm]
$\HTNum$ & Scalar sum of the magnitudes of the momenta of the $H\rightarrow \ell\nu\ell\nu$ and $H\rightarrow bb$ systems, \\
&  \multicolumn{1}{c}{$\HTNum = |\ptmiss +  \ptvec{\ell}{,0} + \ptvec{\ell}{,1}| + |\ptvec{b}{,0} + \ptvec{b}{,1}|$} \\ [0.1cm]
$\HTRatio$ & Ratio of \HTNum and scalar sum of the transverse momenta of the $H$ decay products, \\
& \multicolumn{1}{c}{$\HTRatio = \HTNum / (\met + |\ptvec{\ell}{,0}| + | \ptvec{\ell}{,1}| + |    \ptvec{b}{,0}| + |\ptvec{b}{,1}|)$,} \\
& where \ptvec{\ell(b)\,}{,0 \{1\}} are the transverse momenta of the leading \{subleading\} lepton ($b$-tagged jet) \\ [0.1cm]
\hline
\bottomrule
\end{tabularx}
\end{footnotesize}
\end{center}
\end{table}
% End of text imported from the .//tables/tab_nn_inputs.tex input file
% End of text imported from the .//sections/deep_learning_approach.tex input file
% The next lines are included from the .//sections/signal_selection.tex input file
\subsection{Signal selection criteria}
\label{sec:signal_selection}
 
To define signal selection criteria, the analysis relies on the invariant mass of the two leptons, $m_{\ell\ell}$, and the invariant mass of the two leading
(\pT-ordered) $b$-tagged jets, $m_{bb}$.
Due to spin-correlation effects present in the $H \rightarrow WW^* \rightarrow \ell\nu\ell\nu$ decay within the dominant $\HH \rightarrow \bbWW$ signal
process, the signal events exhibit values of $m_{\ell\ell}$ that are typically below $60$ GeV.
By selecting low values of $m_{\ell\ell}$, the signal purity can therefore be enhanced while rejecting a large
component of the SM $Z$ boson and Top backgrounds.
Additionally, $m_{bb}$ has a peak at the mass of the Higgs boson for the signal process and therefore provides
an effective means to define selections in which the \HH contribution is enhanced.
The signal selection criteria therefore require $m_{\ell\ell} \in (20,60)$ GeV and
$m_{bb} \in (110,140)$ GeV. The $m_{\ell\ell}>20$ GeV requirement is enforced in order to remove contamination
from low-mass resonances and $Z/\gamma^*$ processes.
The signal selection criteria are further broken down into same-flavour (SF), i.e. $ee$ or $\mu\mu$, or different-flavour (DF), i.e.
$e\mu$, regions.
Separating by dilepton flavour enhances the separation power between the signal and
$Z/\gamma^*$ background; the former has roughly equal probabilities for the SF and DF final
states and the latter leads predominantly to SF final states.
 
In addition to the \mll and \mbb requirements, the same-flavour and different-flavour
signal regions, SR-SF and SR-DF, respectively, are defined by requiring high values
of $\dhh$ and are presented in Table \ref{tab:region_defs_yields}.
The chosen threshold values of $\dhh>5.45$ ($5.55$) for SR-SF (SR-DF) are found
to maximise the expected sensitivity to the non-resonant \HH process.
The predicted $\HH \rightarrow \bblvlv$ signal yields in SR-SF and SR-DF are
shown in Table \ref{tab:region_defs_yields}, and are composed of $90\%$ $\HH \rightarrow \bbWW$, $9\%$
$\HH \rightarrow \bbtt$, and $1\%$ $\HH \rightarrow \bbZZ$. The predominance of the $\HH \rightarrow \bbWW$
process over the other two is a result of both its larger overall BF and of the classifier having been trained only on this component of the signal.
% End of text imported from the .//sections/signal_selection.tex input file
% The next lines are included from the .//sections/background_estimation.tex input file
\subsection{Background estimation}
\label{sec:background_estimation}
 
As mentioned in Section \ref{sec:nn_desc}, the dominant backgrounds
expected to contaminate the signal regions are the \bkgtop and $Z/\gamma^*$ processes, specifically
$Z/\gamma^*$ production in association with jets originating from heavy-flavour hadrons ($bb$, $bc$, or $cc$),
subsequently referred to as \bkgzhf.
Subdominant SM processes contribute via $\ttbar$ production in association with an electroweak
vector boson, single Higgs boson production (predominantly via the $\ttbar H$ mode),
$Z/\gamma^*$ production in association with light-flavour jets, and electroweak diboson processes.
There is additionally a minor contribution of background events from non-prompt leptons
produced in semileptonic decays of heavy-flavour hadrons and from misidentified electron
candidates arising from photon conversions and jets.
This background is estimated using events with a same-charge lepton pair, following procedures
described in Ref. \cite{TOPQ-2015-09}, after subtracting the prompt-lepton contribution.
The rest of the SM background processes detailed in Table \ref{tab:data_mc} are estimated primarily using simulation.
 
Dedicated control regions are defined to derive data-driven normalisation corrections for the dominant
background processes: CR-Top for \bkgtop and CR-Z+HF for \bkgzhf.
These normalisation corrections have a uniform prior and are checked in two validation regions, VR-1 and VR-2,
enriched with events from the \bkgtop and \bkgzhf processes.
The control and validation regions are defined
in Table \ref{tab:region_defs_yields} and are kinematically close to the signal regions.
CR-Top (CR-Z+HF) and VR-1 (VR-2) are defined by inverting the $m_{bb}$ ($m_{\ell\ell}$) requirements relative to those of the signal regions but retain a selection of the high \dhh region similar to the signal regions. The \dhh selections were relaxed to increase statistical power, independent checks showed that this did not have a significant impact on the post-fit normalisation corrections in Table \ref{tab:region_defs_yields}.
 
VR-1 keeps only those events with $m_{bb} > 140$ GeV, excluding the region $m_{bb} < 100$ GeV which is included in CR-Top,
due to significant
contamination of \bkgzhf events.
The correlations between the $m_{\ell\ell}$ and $m_{bb}$ observables and \dhh after the preselection are observed
to be small and do not prevent the use of the former two in the construction
of the analysis regions defined in Table \ref{tab:region_defs_yields}, as \dhh is found
to rely mainly on the information provided by the additional input observables listed in Table \ref{tab:nn_inputs}.
This absence of strong correlation ensures that the measurements made in the tails of \dhh
in the control regions can be extrapolated to those in the signal regions.
 
The \bkgtop background in the signal regions
is expected to be composed of approximately equal contributions from the \ttbar and single-top-quark \wt process
and therefore susceptible to the interference effects as described in Section \ref{sec:datamc}.
For this reason, CR-Top and the validation regions are defined so that they have predicted \ttbar and \wt compositions similar to that of the
signal regions.
This ensures that the normalisation correction determined in the fit for the \bkgtop background
results in an accurate estimate of the combined \ttbar and \wt process in the signal regions,
accounting for potential interference effects present in data but not necessarily modelled in MC simulation.
Table \ref{tab:region_defs_yields} compares the observed and predicted event yields,
where the background event yields obtained after background-only fits in the corresponding control regions are also shown.
The post-fit normalisation correction factors for the \bkgtop and \bkgzhf background processes, respectively
$\mu_{\bkgtop} = 0.79 \pm 0.10$ and $\mu_{\bkgzhf} = 1.36 \pm 0.07$, are also shown in Table \ref{tab:region_defs_yields}.
The uncertainties in $\mu_{\bkgtop}$ and $\mu_{\bkgzhf}$ take into account the statistical and systematic
uncertainties due to the experimental sources, as described in Section \ref{sec:systematics}.
 
Distributions of $\dhh$ in the control regions after performing background-only fits to data
in the control regions and applying the \bkgtop and \bkgzhf normalisation corrections are shown in Figure \ref{fig:cr_dhh}.
In the control and validation regions, good agreement between the data and SM prediction provided by the post-fit MC simulation
is observed for the observables relevant to the analysis.
 
% The next lines are included from the .//tables/tab_region_defs_yields.tex input file
 
\begin{table}[!htb]
\begin{center}
\caption{
Analysis region and background estimation summary.
Shown are the definitions of the control, validation, and signal regions used in the
analysis as well as the predicted and observed event yields in each of these
regions.
The predicted yields are shown after background-only
fits to data in the control regions.
The \bkgtop and \bkgzhf post-fit normalisation factors, obtained from background-only fits in
the corresponding control regions, are shown at the bottom of the table.
Also shown is the predicted $\hh \rightarrow \bblvlv$ signal yield in each of the regions,
multiplied by a factor of 20.
Of the $\hh$ yield in the signal regions,
90\% comes from the $\hh \rightarrow bbWW^*$ process,
9\% from the $\hh \rightarrow bb\tau\tau$ process,
and 1\% from the $\hh \rightarrow bbZZ^*$ process.
The uncertainties in each yield and in the normalisation correction
factors account for the statistical and systematic uncertainties
described in Section \ref{sec:systematics}, with those
on the normalisation corrections due only to experimental sources.
}
\label{tab:region_defs_yields}
\begin{footnotesize}
\begin{tabularx}{0.91\textwidth}{l c | c c c c c c}
\toprule
\hline
\multicolumn{8}{c}{\textbf{Region Definitions}} \\
\hline
\multicolumn{2}{l|}{\textbf{Observable}} & \textbf{CR-Top} & \textbf{VR-1} & \textbf{CR-Z+HF} & \textbf{VR-2} & \textbf{SR-SF} & \textbf{SR-DF} \\
\hline
\multicolumn{2}{l|}{Dilepton Flavour} & DF & SF & DF or SF & SF & SF & DF \\
\multicolumn{2}{l|}{$m_{\ell\ell}$ [GeV]} & $(20,60)$ & $(20,60)$ & $(81.2,101.2)$ & $(71.2,81.2)$ & $(20,60)$ & $(20,60)$ \\
& &               &               &                    & or $(101.2, 115)$ & & \\
\multicolumn{2}{l|}{$m_{bb}$ [GeV]}     & $\not\in(100,140)$ & $>140$  & $(100,140)$ & $(100,140)$ & $(110,140)$ & $(110,140)$ \\
\multicolumn{2}{l|}{$\dhh$}            & $>4.5$ & $>4.5$ & $>0$ & $>0$ & $>5.45$ & $>5.55$ \\
\hline
\hline
\multicolumn{8}{c}{\textbf{Event Yields}} \\
\hline
\multicolumn{2}{l|}{Data}            & 108 & 171 & 852 & 157 & 16 & 9 \\
\hline
\multicolumn{2}{l|}{Total Bkg.}      & $108 \pm 10\phantom{..}$                     & $162 \pm 10\phantom{.}$            & $852 \pm 29\phantom{.}$           & $147 \pm 11$
& $14.9 \pm 2.1$  & $4.9 \pm 1.2$ \\
\hline
\multicolumn{2}{l|}{\bkgtop}       & $\phantom{1.}92 \pm 11\phantom{.0}$            & $\phantom{0}77 \pm 10\phantom{.}$  & $\phantom{.}55 \pm 7\phantom{0}$  & \phantom{..}$71 \pm 10$
& \phantom{..}$4.8 \pm 1.4$  & $3.8 \pm 1.1$ \\
\multicolumn{2}{l|}{\bkgzhf}       & $\phantom{1}3.2  \pm 0.5\phantom{0}$           & $\phantom{0}70 \pm 4\phantom{.0}$  & $686 \pm 33\phantom{.}$           & $60 \pm 4$
& \phantom{..}$7.8 \pm 1.4$  & $0.21 \pm 0.05$ \\
\multicolumn{2}{l|}{Other}         & $13.1 \pm 3.4\phantom{0}$                      & $14.2 \pm 1.9\phantom{.}$          & $110 \pm 13\phantom{.}$           & $15.8 \pm 1.2$
& \phantom{..}$2.3 \pm 0.5$  & $0.9 \pm 0.4$ \\
\hline
\multicolumn{2}{l|}{$\hh$ ($\times 20$)}     & $2.70 \pm 0.25$                      & \hspace{0.08cm}$1.03 \pm 0.22$     & \phantom{.}$1.97 \pm 0.11$        & \phantom{..}$1.22 \pm 0.05$
& \phantom{..}$5.0 \pm 0.6$  & $4.8 \pm 0.8$ \\
\hline
\multicolumn{8}{c}{\textbf{Post-fit Normalisation}} \\
\hline
\multicolumn{4}{c|}{$\mu_{\bkgtop} = 0.79 \pm 0.10$} & \multicolumn{4}{c}{ $\mu_{\bkgzhf} = 1.36 \pm 0.07$} \\
\hline
\bottomrule
\end{tabularx}
\end{footnotesize}
\end{center}
\end{table}
% End of text imported from the .//tables/tab_region_defs_yields.tex input file
 
\begin{figure}[!htb]
\begin{center}
\includegraphics[scale=0.40]{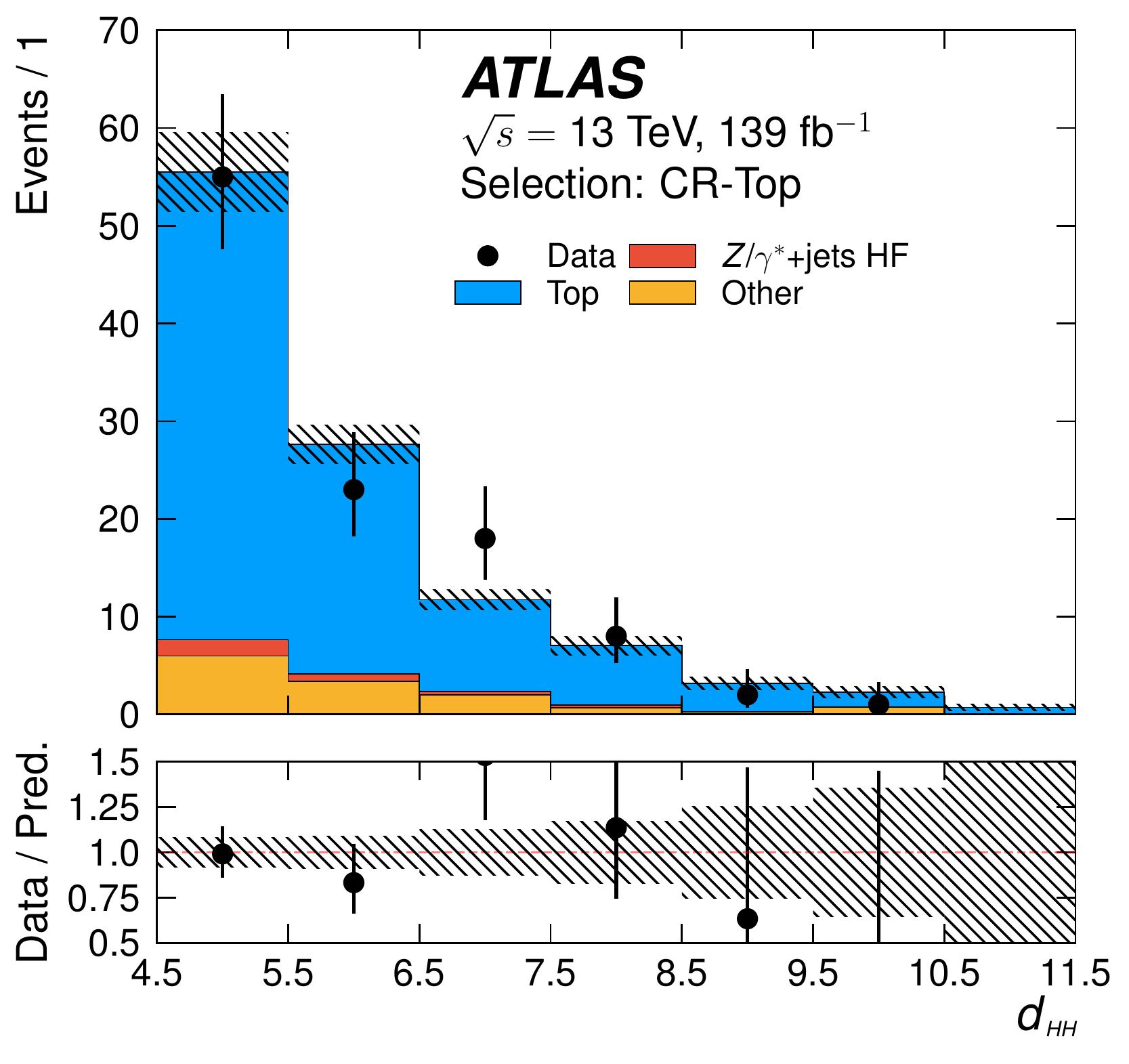}
\includegraphics[scale=0.40]{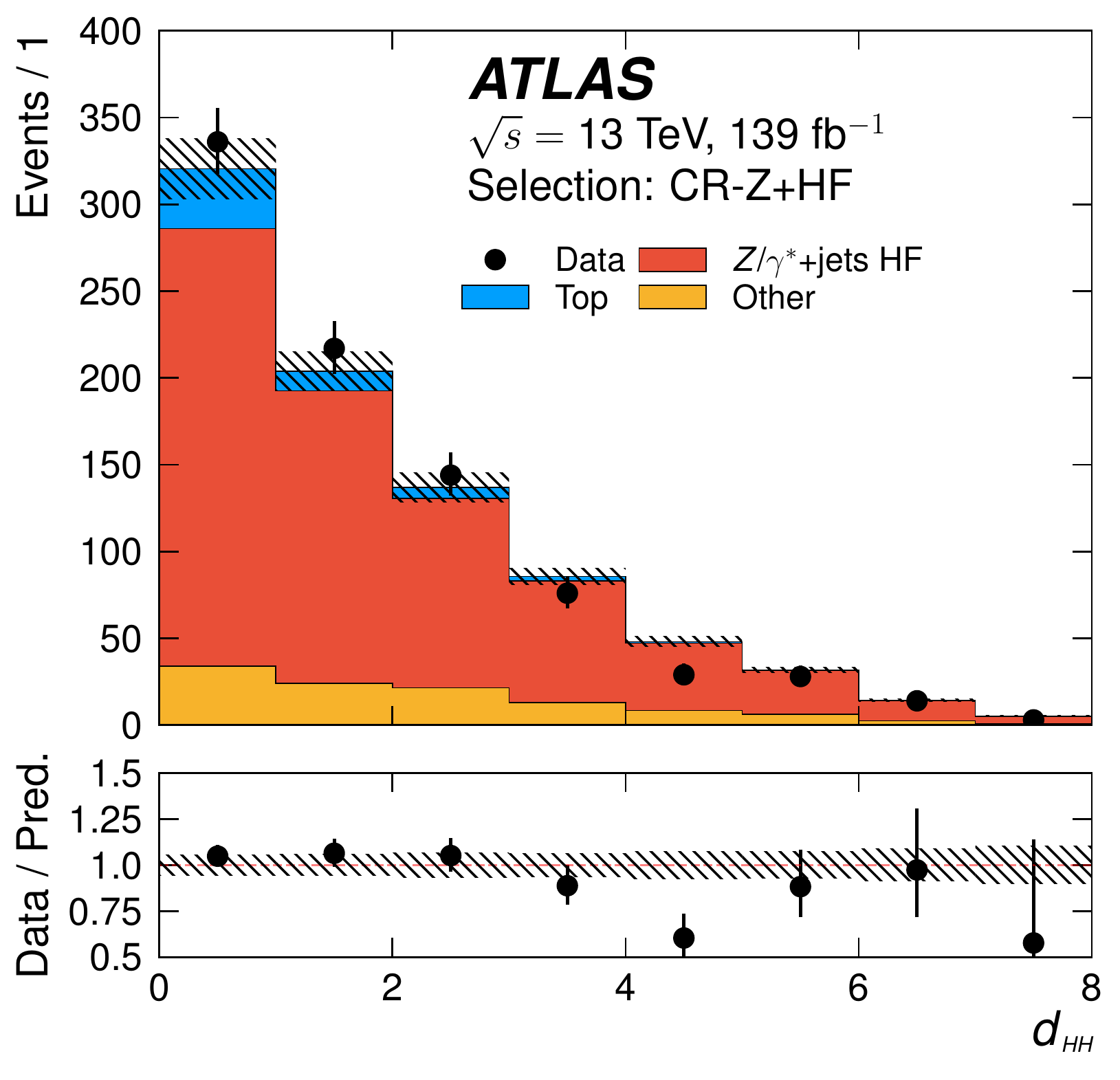}
\caption{
Distributions of $\dhh$ in CR-Top (left) and CR-Z+HF (right).
Distributions are shown after the fit to data under the background-only hypothesis in the shown control regions.
The ratio of the data to the sum of the backgrounds is shown in the lower panel.
The hatched bands indicate the combined statistical and systematic uncertainty.
}
\label{fig:cr_dhh}
\end{center}
\end{figure}
 
% End of text imported from the .//sections/background_estimation.tex input file
% End of text imported from the .//sections/analysis.tex input file
% The next lines are included from the .//sections/systematics.tex input file
\section{Systematic uncertainties}
\label{sec:systematics}
 
The analysis evaluates several sources of systematic uncertainty for the signal and background processes,
which are classified as either \textit{experimental} (detector or luminosity related) or theoretical
modelling uncertainties.
Statistical uncertainties of the simulated event samples are also taken into account.
The main uncertainty components are summarised in Table \ref{tab:syst}.
MC modelling
uncertainties in the \bkgtop and \bkgzhf background estimates are dominant, followed
by statistical and detector uncertainties.
 
The normalisation corrections of the \bkgtop and \bkgzhf background processes are determined
primarily by the data events in the control regions when performing the statistical analysis.
These corrections take into account the statistical and systematic uncertainties due to the
experimental sources, as described later in this section.
In addition, the systematic uncertainties in the theoretical modelling of these processes are
applied as uncertainties in the corrected predictions in the signal regions using the
following procedures.
The uncertainties in the estimated \bkgtop background event yields due to parton shower modelling are assessed as the
difference between the predictions of \powhegbox\ showered with \pythia\ or \herwig, and
those due to the choice of event generator are assessed by comparing the predictions of \powhegbox\
or \mgfive~\cite{ATL-PHYS-PUB-2016-004}, both showered with \pythia. The uncertainties due to missing higher-order
corrections are estimated by changing the renormalisation and factorisation scales (\muR and \muF, respectively) up and
down by a factor of two (8-points variation). The uncertainties due to the modelling of initial- and final-state
radiation (ISR and FSR, respectively) in the generators used to simulate the \bkgtop background processes are evaluated using the method described in Ref. \cite{ATL-PHYS-PUB-2016-004}.
The \bkgtop background composition is varied within the uncertainties in the theoretical predictions
for the \ttbar and single-top-quark \wt cross-sections \cite{Czakon:2013goa,ATLAS-CONF-2013-102,Kidonakis:2010ux}.
The uncertainty arising from the interference between
the NLO predictions for \ttbar and \wt processes is estimated by taking the difference between the predicted
\bkgtop background yields obtained with the DR and DS schemes used for the NLO \wt calculation \cite{ATL-PHYS-PUB-2016-004}.
The uncertainties due to PDF variations are computed as the envelope of the central values of the nominal NNPDF3.0 PDF set and the CT14, MMHT14, and PDF4LHC15\_30PDF PDF sets \cite{Butterworth:2015oua}.
All uncertainties except those in the
scale variations, cross-section, and interference are considered as fully correlated between the \ttbar
and \wt processes.
The \bkgzhf modelling uncertainties are estimated using the nominal \sherpa\,2.2.1 samples
by considering different merging (CKKW-L) \cite{Lonnblad:2012ix} and resummation scales. The uncertainties due to PDF variations and changes in \muR and \muF are calculated using the same procedures as for the \bkgtop backgrounds.
An additional uncertainty in the $Z/\gamma^*$ process is computed by taking the difference between the nominal \sherpa\,2.2.1 samples with
samples generated using \mgfive+\pythia8.
The dominant uncertainties in the total background estimates in SR-SF are the \bkgzhf modelling uncertainties ($8\%$), primarily that due to
comparison of \sherpa\,2.2.1 and \mgfive, and the parton shower uncertainty affecting the \bkgtop background process ($5\%$).
The uncertainties in the background estimates in SR-DF are
dominated by the uncertainty due to the parton shower affecting the \bkgtop background process ($12\%$),
the uncertainty in the \bkgtop normalisation correction $\mu_{\bkgtop}$ ($10\%$),
the uncertainty due to the comparison between the generators used for the \bkgtop process ($7.5\%$),
and the uncertainty due to the modelling of ISR and FSR in the \bkgtop process ($5\%$).
 
Systematic uncertainties in the signal acceptance due to varying \muR and \muF, as well as PDF-induced uncertainties,
are evaluated using the same procedure
as for the \bkgtop background process. The resulting scale (PDF) uncertainties are $<3\%$ ($<1\%$)
in both signal regions. The uncertainty due to the parton shower modelling is computed by comparing \herwig7
with \pythia8, and is found to be $8\%$ ($9\%$) in SR-SF (SR-DF).
The uncertainty in the \hh production cross-section, evaluated to be $5\%$, is included as an uncertainty in $\sigma^{\text{\tiny{SM}}}(gg\rightarrow HH)$ when computing the upper limits on the cross-section ratio in Table \ref{tab:xsec_limits}. This value is the quadrature sum of the scale, PDF+$\alpha_s$, and top mass contributions as reported by the LHCXSWG \cite{Grazzini:2018bsd}.
 
The uncertainties due to experimental sources arise primarily from the mismeasurement of reconstructed object
momenta and from the mismodelling of reconstruction efficiencies.
These uncertainties include uncertainties from the mismodelling of the jet energy scale (JES) \cite{PERF-2016-04}
and jet energy resolution (JER) \cite{ATL-PHYS-PUB-2015-015}.
Additional uncertainties for $b$-tagged jets arise from the mismodelling of the $b$-tagging efficiency~\cite{FTAG-2018ttbar}
and from the mismodelling of the rates at which charm- and light-flavoured jets
are selected as $b$-tagged jets~\cite{ATLAS-CONF-2018-001,ATLAS-CONF-2018-006}.
Lepton-related uncertainties arise from the mismodelling of the electron \cite{LatestEgamma} (muon \cite{PERF-2015-10})
reconstructed energy (momentum) measurements, as well as in the mismodelling of their reconstruction and identification efficiencies~\cite{LatestEgamma,PERF-2015-10}.
The $\met$ scale and resolution \cite{PERF-2016-07} uncertainties, as well as uncertainties
from the mismodelling of \pileup, trigger efficiency and luminosity, are also
taken into account.
The uncertainty in the combined 2015--2018 integrated luminosity is $1.7\%$ \cite{ATLAS-CONF-2019-021}, obtained using the LUCID-2 detector \cite{LUCID2}
for the primary luminosity measurements.
The combined effect of the experimental sources of systematic uncertainty in the predicted background
yields is summarised in Table \ref{tab:syst} and is dominated by the JER, with all other contributions found
to be negligible.
 
% The next lines are included from the .//tables/tab_syst.tex input file
\begin{table}[!htb]
\begin{center}
\caption{
Breakdown of the main uncertainty components in the background estimates in the
two signal regions for the \bkgtop, \bkgzhf, and all other (``Other'') backgrounds.
The uncertainty components associated with the total background estimate in the signal
regions (the sum of \bkgtop, \bkgzhf, and Other) is listed under ``Total Bkg.''.
As in the upper half of Table~\ref{tab:region_defs_yields}, all uncertainties are shown
``post-fit''. The percentages show the size of the uncertainty relative
to the expected background in each column and uncertainties can be correlated,
not necessarily adding in quadrature to the total uncertainty in
each column or across each row. Uncertainties in the post-fit normalisation factors,
$\mu_{\bkgtop}$ and $\mu_{\bkgzhf}$, are only applicable for the \bkgtop and
\bkgzhf processes.
}
\label{tab:syst}
\begin{tabular}{ l | r r r r | r r r r}
\toprule
\hline
\textbf{Uncertainty} $[\%]$ & \multicolumn{4}{c|}{\textbf{SR-SF}} & \multicolumn{4}{c}{\textbf{SR-DF}} \\
& \bkgtop & \bkgzhf & Other & Total Bkg. & \bkgtop & \bkgzhf & Other & Total Bkg. \\
\hline
Total uncertainty & $28$ & $18$ & $20$ & $14$ & $30$ & $26$ & $41$ & $25$ \\
\hline
Theoretical & $21$ & $15$ & $17$ & $11$ & $20$ & $15$ & $40$ & $17$ \\
Experimental & $12$ & $<5$ & $8$ & $<5$ & $15$ & $17$ & $8$ & $12$ \\
MC statistics & $8$ & $8$ & $6$ & $8$ & $13$ & $13$ & $7$ & $11$ \\
$\mu_{\bkgtop}$, $\mu_{\bkgzhf}$ & $13$ & $5$ & n/a & $5$ & $13$ & $5$ & n/a & $10$ \\
\hline
\bottomrule
\end{tabular}
\end{center}
\end{table}
% End of text imported from the .//tables/tab_syst.tex input file
% End of text imported from the .//sections/systematics.tex input file
% The next lines are included from the .//sections/results.tex input file
\section{Results}
\label{sec:result}
 
In order to extract information about the $\hh \rightarrow \bblvlv$ signal cross-section, a counting experiment
is performed with a profile-likelihood fit
\cite{Cowan:2010js} simultaneously across the CR-Top, CR-Z+HF, SR-SF, and SR-DF regions using the predicted
and observed event counts in each region as inputs.
The \bkgtop and \bkgzhf normalisation corrections are also extracted from this fit and are found to differ negligibly
from those presented in Table \ref{tab:region_defs_yields}.
All sources of systematic and statistical uncertainty
in the signal and background models are implemented as deviations from the nominal model, scaled by
nuisance parameters that are profiled in the fit.
The $p$-value corresponding to the background-only hypothesis, giving the probability
that the data in the signal regions be at least as incompatible with the background-only
hypothesis as that observed in SR-SF and SR-DF, is $p_0 = 0.15$ and corresponds to $1.05\sigma$ significance.
Distributions of $m_{bb}$, $m_{\ell\ell}$, and $\dhh$ after performing background-only fits to data
in the control regions and applying the \bkgtop and \bkgzhf normalisation corrections
are shown in Figure \ref{fig:sr_dhh}.
The signal selection criteria are imposed on all observables shown in Figure \ref{fig:sr_dhh} apart
from the one being plotted, except that the $\dhh$ requirement for the $m_{bb}$ and $m_{\ell\ell}$ distributions
is relaxed to $\dhh > 5$.
No significant excess of events over the expected SM background is observed and upper limits are set
on non-resonant Higgs boson pair production at $95\%$ confidence level (CL) using the $\text{CL}_{\text{s}}$ method \cite{Read:2002hq}.
Table \ref{tab:xsec_limits} presents these upper limits and comparisons with the SM prediction.
The observed (expected) limit at $95\%$ CL is $1.2$ ($0.9$) pb, corresponding
to $40$ ($29$) times the SM prediction.
 
\begin{figure}[!htb]
\begin{center}
\includegraphics[width=0.32\textwidth]{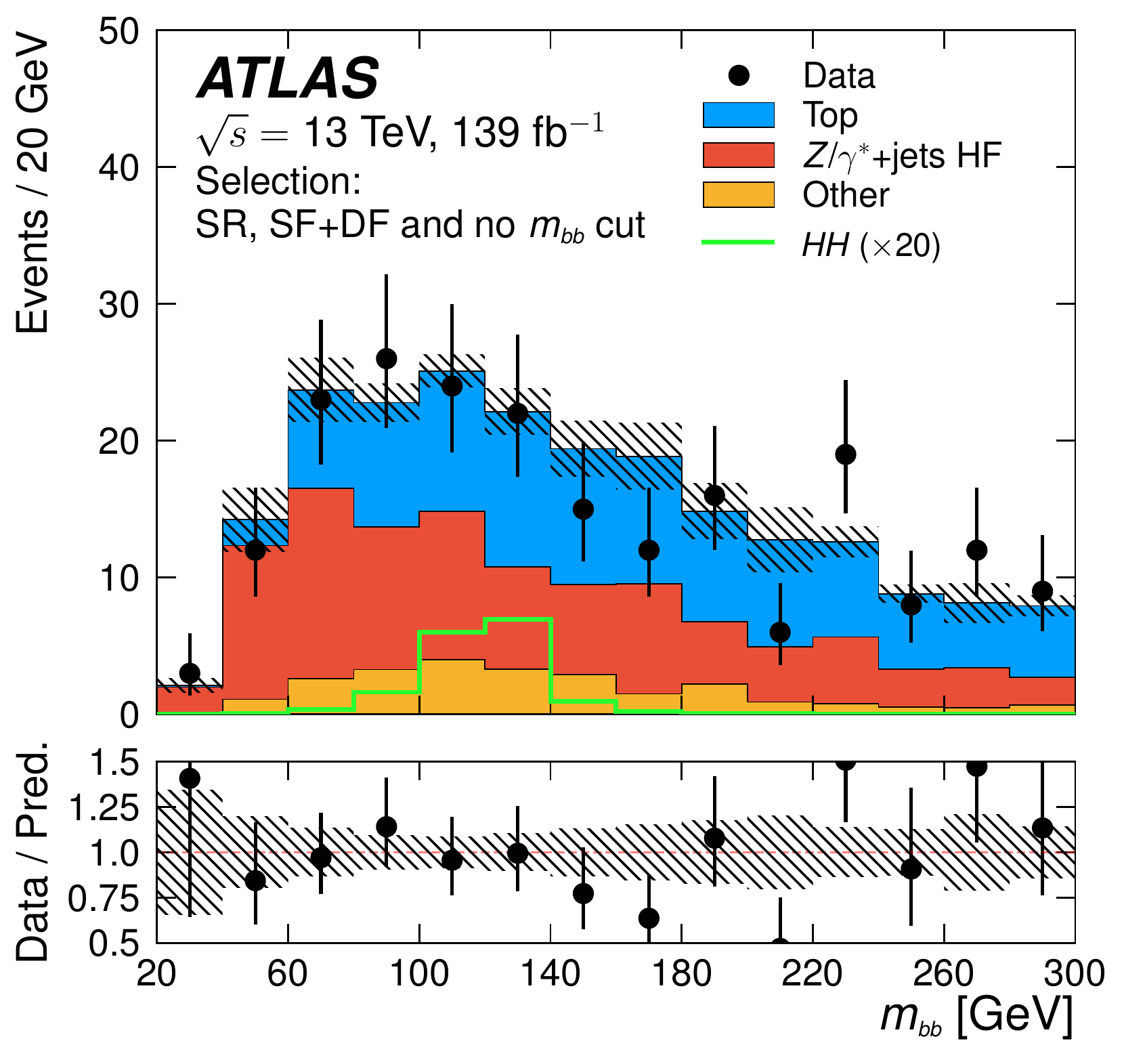}
\includegraphics[width=0.32\textwidth]{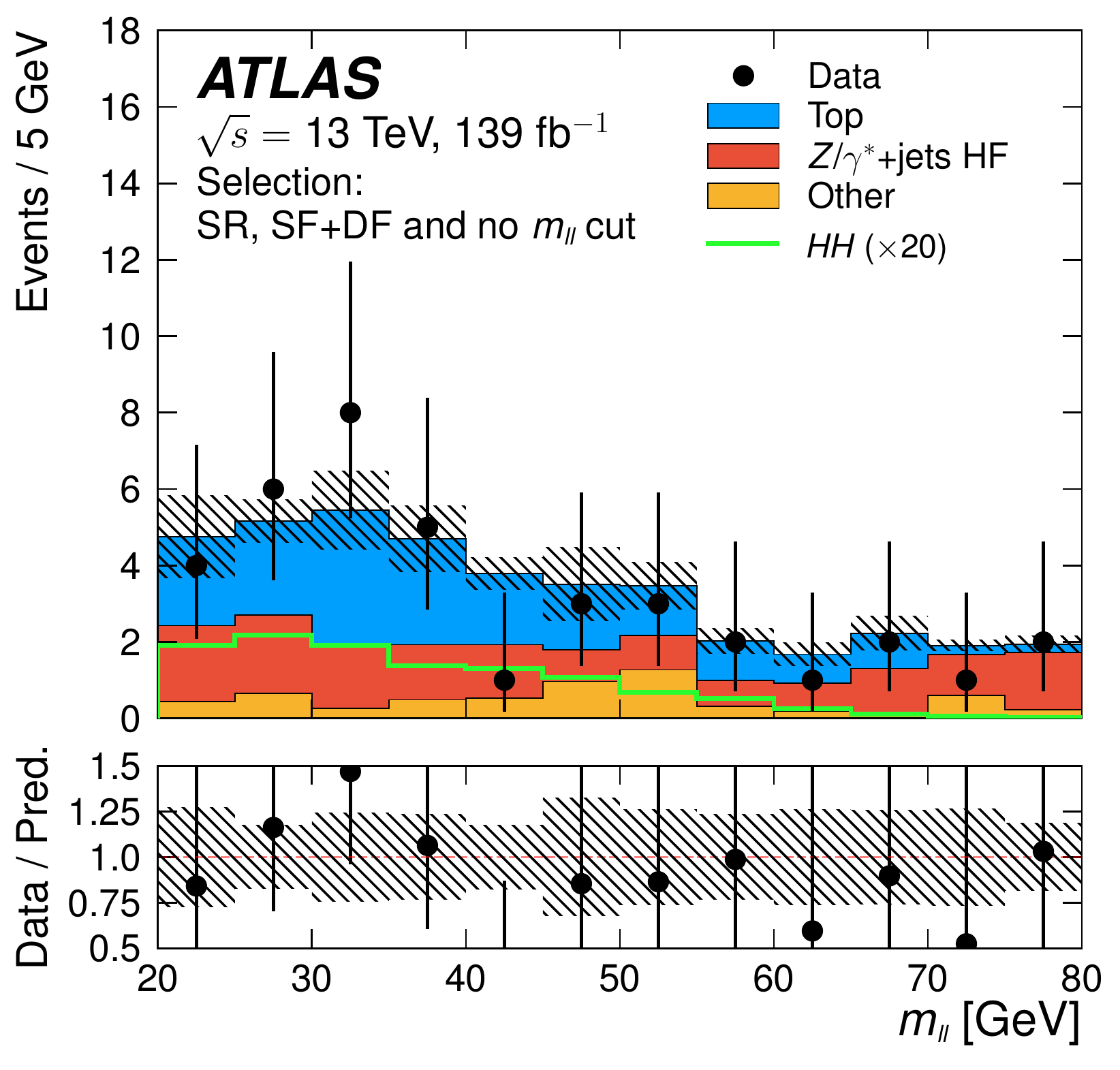}
\includegraphics[width=0.32\textwidth]{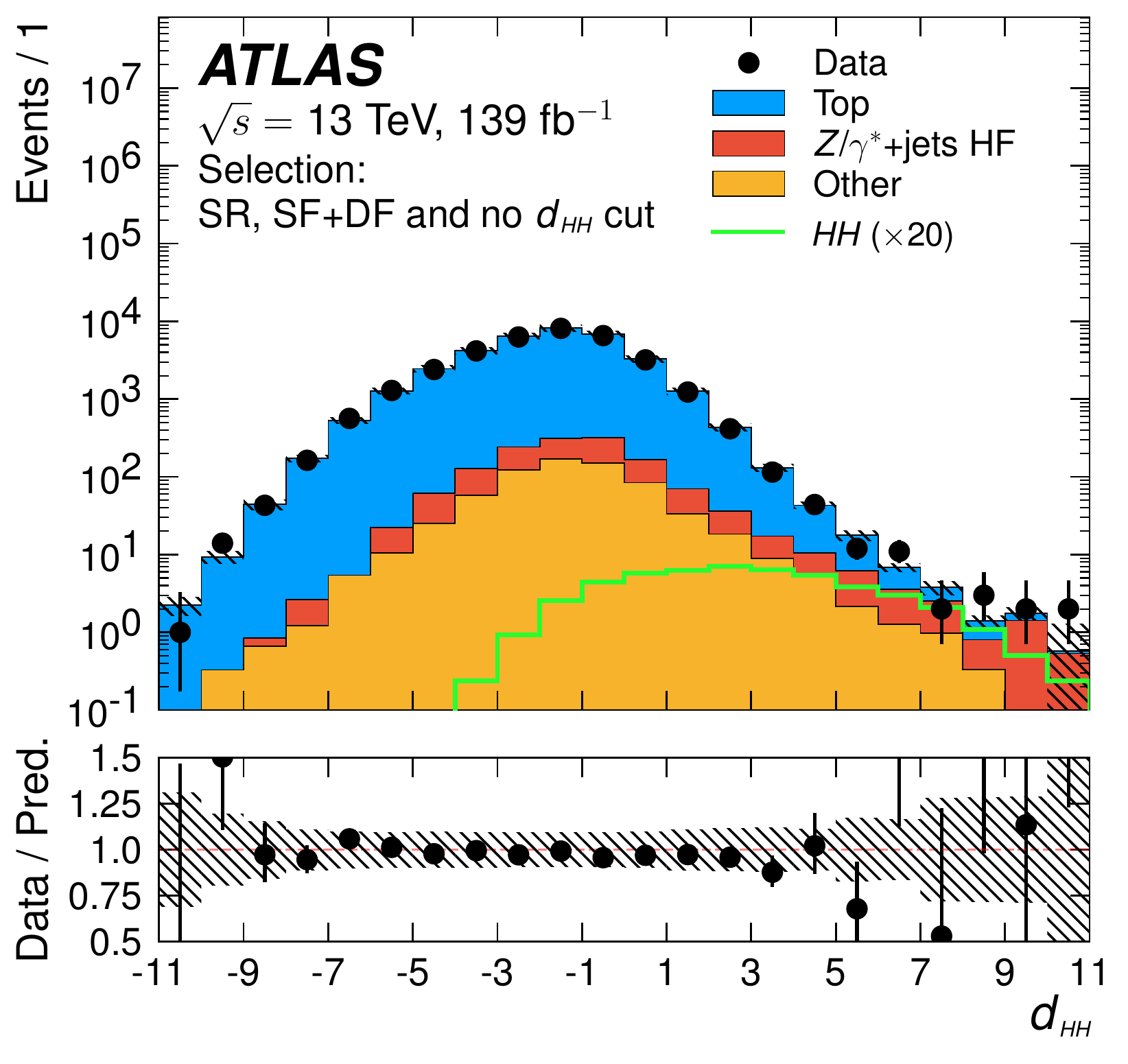}
\caption{
Distributions of $m_{bb}$ (left), $m_{\ell\ell}$ (middle), and the
discriminant $\dhh$ (right). The distributions are shown after the fit to data
in the control regions under the background-only hypothesis.
Each distribution includes both the SF and DF events and
imposes signal selection requirements on all quantities except the one
being plotted, but the requirement on $\dhh$ has been relaxed to
$\dhh>5$ for the distributions of $m_{bb}$ and $m_{\ell\ell}$.
The $\HH \rightarrow \bblvlv$ signal (``\HH'') is overlaid and has its cross-section
scaled by a factor of 20 relative to the SM prediction for visualisation purposes.
The ratio of the data to the sum of the backgrounds is shown in the lower
panel of each figure. The hatched bands indicate the combined statistical and
systematic uncertainty.
}
\label{fig:sr_dhh}
\end{center}
\end{figure}

\begin{table}[!htb]
\begin{center}
\caption{
Observed and expected upper limits on the ggF-initiated non-resonant \hh production cross-section at $95\%$ CL
and their ratios to the SM prediction ($\sigma^{\text{\tiny{SM}}}(gg\rightarrow HH) = 31.05\pm1.90$ fb \cite{deFlorian:2016spz,Dawson:1998py,Borowka:2016ehy,Baglio:2018lrj,deFlorian:2013jea,Shao:2013bz,deFlorian:2015moa,Grazzini:2018bsd}). The $\pm 1 \sigma$
and $\pm 2 \sigma$ variations about the expected limit are also shown. Uncertainties in the SM cross-section are taken into account when computing the upper limits on the cross-section ratio.
}
\begin{tabular}{l | c c c c c c }
\toprule
\hline
& $-2 \sigma$ & $-1\sigma$ & \textbf{Expected} & $+1\sigma$ & $+2\sigma$ & \textbf{Observed} \\
\hline
$\sigma \left( gg \rightarrow \hh \right)$ [pb] & 0.5  & 0.6  & 0.9  & 1.3  & 1.9  & 1.2 \\ 
$\sigma \left( gg \rightarrow \hh \right) / \sigma^{\mbox{\tiny{SM}}} \left( gg \rightarrow \hh \right)$      & 14 & 20 & 29 & 43 & 62 & 40\\ 
\hline
\bottomrule
\end{tabular}
\label{tab:xsec_limits}
\end{center}
\end{table}
% End of text imported from the .//sections/results.tex input file
 
\FloatBarrier
% The next lines are included from the .//sections/conclusion.tex input file
\section{Conclusions}
\label{sec:conclusion}
 
A search for non-resonant Higgs boson pair production, as predicted by the SM, is presented in the final state with at least two $b$-tagged jets and exactly two leptons with opposite electric
charge, where one of the Higgs bosons decays to $bb$ and the other decays to either $WW^*$, $ZZ^*$, or $\tau\tau$.
The analysis uses \pp collision data recorded at $\sqrt{s} = 13$ TeV by the ATLAS
detector at the LHC, corresponding to an integrated luminosity of 139 fb$^{-1}$.
The data are in agreement with the predictions for the SM background processes.
An observed (expected) $95\%$ CL upper limit is set on the cross-section for the production
of Higgs boson pairs, corresponding to $40$ ($29$) times the SM prediction.
These limits are comparable to the previous leading searches for non-resonant Higgs boson pair production
performed by the ATLAS and CMS experiments.
% End of text imported from the .//sections/conclusion.tex input file

\section*{Acknowledgements}
 
% The next lines are included from the .//acknowledgements/Acknowledgements.tex input file
 
We thank CERN for the very successful operation of the LHC, as well as the
support staff from our institutions without whom ATLAS could not be
operated efficiently.
 
We acknowledge the support of ANPCyT, Argentina; YerPhI, Armenia; ARC, Australia; BMWFW and FWF, Austria; ANAS, Azerbaijan; SSTC, Belarus; CNPq and FAPESP, Brazil; NSERC, NRC and CFI, Canada; CERN; CONICYT, Chile; CAS, MOST and NSFC, China; COLCIENCIAS, Colombia; MSMT CR, MPO CR and VSC CR, Czech Republic; DNRF and DNSRC, Denmark; IN2P3-CNRS, CEA-DRF/IRFU, France; SRNSFG, Georgia; BMBF, HGF, and MPG, Germany; GSRT, Greece; RGC, Hong Kong SAR, China; ISF and Benoziyo Center, Israel; INFN, Italy; MEXT and JSPS, Japan; CNRST, Morocco; NWO, Netherlands; RCN, Norway; MNiSW and NCN, Poland; FCT, Portugal; MNE/IFA, Romania; MES of Russia and NRC KI, Russian Federation; JINR; MESTD, Serbia; MSSR, Slovakia; ARRS and MIZ\v{S}, Slovenia; DST/NRF, South Africa; MINECO, Spain; SRC and Wallenberg Foundation, Sweden; SERI, SNSF and Cantons of Bern and Geneva, Switzerland; MOST, Taiwan; TAEK, Turkey; STFC, United Kingdom; DOE and NSF, United States of America. In addition, individual groups and members have received support from BCKDF, CANARIE, CRC and Compute Canada, Canada; COST, ERC, ERDF, Horizon 2020, and Marie Sk{\l}odowska-Curie Actions, European Union; Investissements d' Avenir Labex and Idex, ANR, France; DFG and AvH Foundation, Germany; Herakleitos, Thales and Aristeia programmes co-financed by EU-ESF and the Greek NSRF, Greece; BSF-NSF and GIF, Israel; CERCA Programme Generalitat de Catalunya, Spain; The Royal Society and Leverhulme Trust, United Kingdom.
 
The crucial computing support from all WLCG partners is acknowledged gratefully, in particular from CERN, the ATLAS Tier-1 facilities at TRIUMF (Canada), NDGF (Denmark, Norway, Sweden), CC-IN2P3 (France), KIT/GridKA (Germany), INFN-CNAF (Italy), NL-T1 (Netherlands), PIC (Spain), ASGC (Taiwan), RAL (UK) and BNL (USA), the Tier-2 facilities worldwide and large non-WLCG resource providers. Major contributors of computing resources are listed in Ref.~\cite{ATL-GEN-PUB-2016-002}.
 
% End of text imported from the .//acknowledgements/Acknowledgements.tex input file

\clearpage

\printbibliography
 
\clearpage \input{atlas_authlist}

\end{document}

%% file: atlas_authlist.tex
% ATLAS Collaboration author list
% Reference date of HDBS-2018-33 is 2019-05-31
% Author list last updated on date 04-DEC-19
% Data extracted on 04-Dec-2019 for paper reference HDBS-2018-33
% at 9:59pm
 
\begin{flushleft}
{\Large The ATLAS Collaboration}

\bigskip

G.~Aad$^\textrm{\scriptsize 101}$,    
B.~Abbott$^\textrm{\scriptsize 128}$,    
D.C.~Abbott$^\textrm{\scriptsize 102}$,    
A.~Abed~Abud$^\textrm{\scriptsize 70a,70b}$,    
K.~Abeling$^\textrm{\scriptsize 53}$,    
D.K.~Abhayasinghe$^\textrm{\scriptsize 93}$,    
S.H.~Abidi$^\textrm{\scriptsize 167}$,    
O.S.~AbouZeid$^\textrm{\scriptsize 40}$,    
N.L.~Abraham$^\textrm{\scriptsize 156}$,    
H.~Abramowicz$^\textrm{\scriptsize 161}$
% ...
\bigskip
\\

% ...
$^{40}$Niels Bohr Institute, University of Copenhagen, Copenhagen; Denmark.\\
% ...
$^{53}$II. Physikalisches Institut, Georg-August-Universit\"{a}t G\"ottingen, G\"ottingen; Germany.\\
% ...
$^{70}$$^{(a)}$INFN Sezione di Pavia;$^{(b)}$Dipartimento di Fisica, Universit\`a di Pavia, Pavia; Italy.\\
% ...
$^{93}$Department of Physics, Royal Holloway University of London, Egham; United Kingdom.\\
% ...
$^{101}$CPPM, Aix-Marseille Universit\'e, CNRS/IN2P3, Marseille; France.\\
$^{102}$Department of Physics, University of Massachusetts, Amherst MA; United States of America.\\
% ...
$^{128}$Homer L. Dodge Department of Physics and Astronomy, University of Oklahoma, Norman OK; United States of America.\\
% ...
$^{156}$Department of Physics and Astronomy, University of Sussex, Brighton; United Kingdom.\\
% ...
$^{161}$Raymond and Beverly Sackler School of Physics and Astronomy, Tel Aviv University, Tel Aviv; Israel.\\
% ...
$^{167}$Department of Physics, University of Toronto, Toronto ON; Canada.\\
% ...
$^{a}$ Also at Borough of Manhattan Community College, City University of New York, New York NY; United States of America.\\
$^{b}$ Also at CERN, Geneva; Switzerland.\\
% ...

\end{flushleft}

% Created with Glance <Atlas.Glance@cern.ch>